\documentclass{article}

\usepackage{amsmath}
\usepackage{amssymb}
\usepackage{graphicx}
\usepackage{textcomp}
\usepackage{hyperref}
\usepackage{subfigure}
\usepackage{authblk}
\usepackage[a4paper, total={6.6in, 9in}]{geometry}

\begin{document}
\title{Accurate simulation of the European XFEL scintillating screens point spread function}
\author{A. Novokshonov}
\affil{Deutsches Elektronen-Synchrotron DESY, Hamburg, Germany}
\date{}
\maketitle

\begin{abstract}
The European XFEL is equipped with scintillating screens as a profile measurement monitor. The scintillating material used is Gadolinium Aluminium Gallium Garnet doped with Cerium (GAGG:Ce). At most of the stations, the screen is positioned perpendicular to the electron beam, with scintillation observed at a backward angle. The scintillator thickness is usually $200 \, \mu m$, making the resolution worse in the plane with the angle, as it allows for the entire particle track within the scintillator to be seen. Besides, aberrations are introduced by the objective used. This study outlines an accurate simulation of the point spread function (PSF) caused by all distortions of the optical system and, in addition, a method to improve the screens resolution by including the PSF into a fitting function, assuming a Gaussian beam shape.
\end{abstract}

\section*{Introduction}
The European XFEL employs scintillating screens rather than more common optical transition radiation (OTR) monitors, as the latter exhibit coherent effects induced by micro-bunching instabilities, which can spoil the image quality and hinder accurate beam size resolution~\cite{Wesch_COTR}. The scintillating screens are not susceptible to these effects. But their resolution is comparatively lower. This trade-off, in principle, is acceptable to avoid the image degradation associated with OTR monitors. And even with the lower resolution the scintillating screens meet the requirements of the European XFEL. However, the resolution may always be improved by knowing the screens PSF, which can be later taken used in an image treatment.

\begin{figure}[!hbt]
	\centering
	\includegraphics[width=0.3\textwidth]{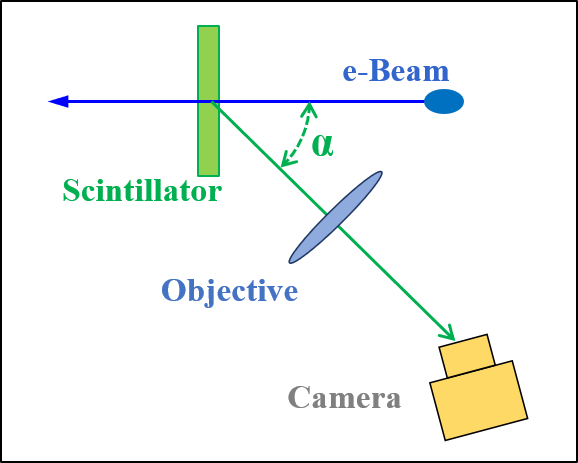}
	\caption{European XFEL scintillating station scheme.}
	\label{fig:stations_schemes}
\end{figure}

A schematic of the scintillating monitor optics is illustrated in Fig.\ref{fig:stations_schemes}. Two objectives, \textit{Schneider-Kreuznach Makro-Symmar 5.6/180} and \textit{5.9/120}, facilitate magnifications of $1:1$ and $1:2$, respectively, to accommodate different resolution needs. The detailed description of the stations geometry may be found in~\cite{Wiebers_Station_description}. The scintillator employed is a $200$ $\mu m$ thick GAGG:Ce. However, this thickness compromises the monitor's resolution in the observation angle's plane. And there are two possible ways of improvement. The first involves using a thinner scintillator, though this reduces light output and increases the material's fragility, making it more challenging to handle. The second solution proposes reducing the observation angle, although physical and operational constraints limit this approach, including the impossibility of positioning the setup directly on the beam axis and potential space restrictions. There is in general another solution - to put a $45^\circ$ (for example) mirror after the scintillator. This would solve most of the problems, but at the electron energies (up to $17\,\, GeV$) and the bunch charge (up to $1\,\, nC$) the mirror's degradation or damage are inevitably.

If it's not possible to entirely eliminate the impact of the observation angle, it becomes essential to accurately determine the point spread function (PSF) of the system. The PSF can either be directly measured or theoretically calculated. However, the measurement presents challenges; it can be technically demanding and the system's geometry might alter over time, necessitating repeated measurements of the PSF. Fortunately, there exists a practical alternative: the entire setup can be modeled in Ansys Zemax Opticstudio\textsuperscript{\textcopyright}~\cite{OpticStudio}. This approach not only circumvents the complexities associated with direct measurement but also allows for flexible adaptation to any changes in system geometry without the need for continuous empirical recalibration.

\section*{Image formation}
This section delves into the mechanism of image formation within the optical system. At its core, the PSF serves as the optical system's response characteristic~\cite{Optics}. It represents the image produced by an infinitely small point source. Thus, the formation of any image can be conceptualized as a convolution of the original source with the PSF:
\begin{equation}
	\label{eq:image_formation1}
	Image = Source * PSF.
\end{equation}
The conventional PSF encapsulates all distortions introduced by the optical system - various types of optical aberrations. However, this formula is only valid for the plane without observation angle. The other plane introduces an additional layer of distortion - a geometrical one, caused exactly by the scintillator thickness and the observation angle. Consequently, the image in this plane incorporates both aberrational and geometrical distortions and mathematically defined by:
\begin{equation}
	\label{eq:image_formation2}
	Image = Source * PSF_{aber} * PSF_{scint},
\end{equation}
where $PSF_{aber}$ accounts for aberrational distortions and $PSF_{scint}$ for the scintillator geometrical distortions. Understanding both PSFs is crucial for accurately restoring the source image through deconvolution. Deconvolution, by nature, is a complex and often challenging process~\cite{Chris_Solomon}, especially when applied to standard profile monitors like those utilizing scintillating screens. However, the task is considerably simplified if the source shape can be reasonably approximated in advance. In our case, we hypothesize that the source exhibits a Gaussian shape, leading to the following representation of the image formation process:
\begin{equation}
	\label{eq:image_formation3}
	Image = Gaussian * PSF_{aber} * PSF_{scint}.
\end{equation}

However, it's important to emphasize that knowledge of the PSFs is imperative, irrespective of our ability to approximate the source's shape. The PSFs fundamentally determine the system's resolution and are essential for any attempt at image restoration, highlighting the significance of their accurate characterization.

As it was mentioned, the $PSF_{scint}$ only appears in the observation angle plane. Hence the other plane is only affected by the $PSF_{sber}$.

\section*{Modeling}
Both PSFs were modeled in OpticStudio\textsuperscript{\textcopyright}, employing two distinct modes for accuracy: sequential and non-sequential. A detailed description of the modes is available in~\cite{OpticStudio}. We only mention here that optical rays in sequential mode only hit each surface ones and travel sequentially from surface 1 to surface 2 and so on. Whereas the non-sequential mode allows the rays to hit the same surface many times. Based on that the $PSF_{aber}$ was simulated using the sequential mode, and the $PSF_{scint}$ in the non-sequential one. This approach ensures that each PSF's unique characteristics are accurately captured and modeled, reflecting the complexities inherent in the optical system.

\subsection*{1:2 optics}
\begin{figure}[!hbt]
	\centering
	\begin{minipage}[c]{0.25\textwidth}
		\centering
		\includegraphics[width=0.95\textwidth]{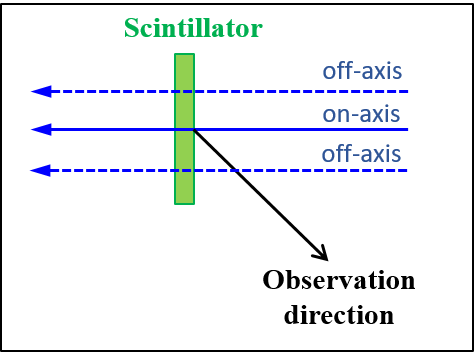}
		\caption{On-axis (solid line) and off-axis (dashed lines) electrons positions.}
		\label{fig:on_off_axis}
	\end{minipage}
	\hspace*{0.5cm}
	\begin{minipage}[c]{0.65\textwidth}
		\centering
		\includegraphics[width=0.9\textwidth]{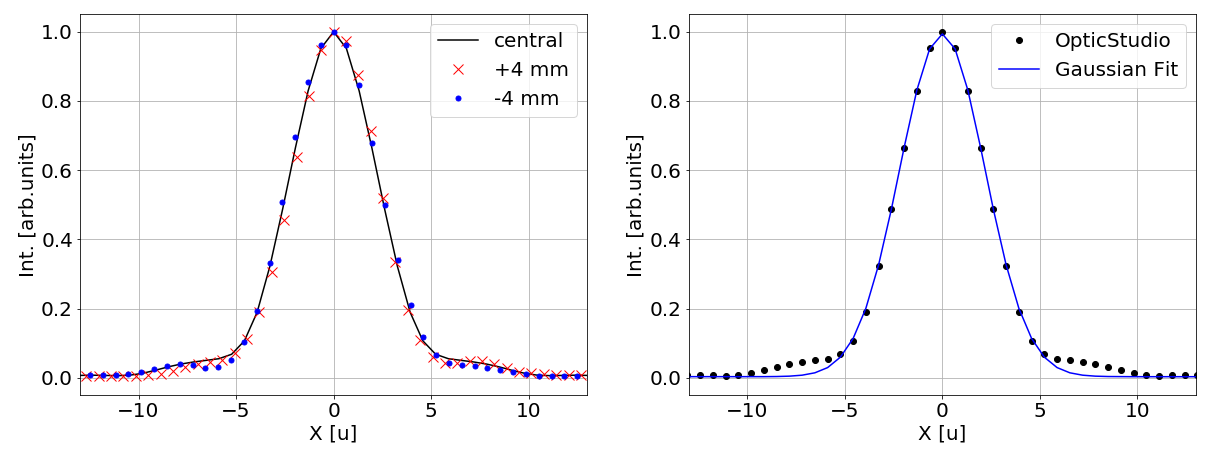}
		\caption{Results of $PSF_{aber}$ simulation: left - comparison of on--axis and off--axis PSFs; right - 			fitting of the central one.}
		\label{fig:psf_aber}
	\end{minipage}
\end{figure}

First lets consider the $1:2$ magnification station, utilizing the \textit{Schneider-Kreuznach Makro-Symmar 5.9/120} objective. In this setup, the distances from the source to the objective and from the objective to the image were precisely set at $360$ mm and $180$ mm, respectively. Additionally, the camera was tilted at an angle of $26.5^\circ$ to accommodate the Scheimpflug principle, ensuring that defocusing effects are correctly mitigated. There were three PSFs simulated - on-axis and two off-axis ones. A schematic depiction of what these are is shown in Fig.~\ref{fig:on_off_axis}. Although the Scheimpflug geometry is used to mitigate any defocusing caused by the observation tilt, three different PSFs have been considered to make sure. The outcome of the simulation is illustrated in Fig.~\ref{fig:psf_aber}. The left part of the figure displays both on-axis and off-axis PSFs with the latter adjusted to the center for comparative purposes to demonstrate the correction of defocusing by the Scheimpflug geometry. The off-axis PSFs are located at $\pm 4$ mm from the screen's center, and the fact they match the central one proves the Scheimpflug correction. Otherwise the offaxis one were broader than the central. The right figure presents a Gaussian fitting of the central PSF, yielding a fit result of $\sigma_{aber}=2.18\, \mu m$, indicating the system's high-resolution capability.

\begin{figure}[!hbt]
	\centering
	\begin{minipage}[c]{0.25\textwidth}
		\centering
		\includegraphics[width=0.95\textwidth]{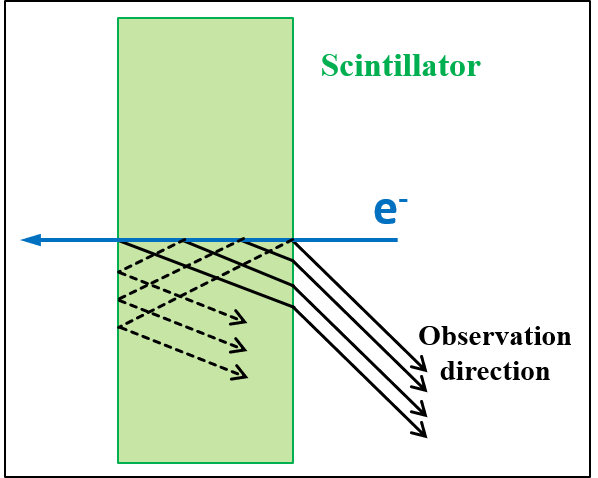}
		\caption{Non-sequential simulation scheme.}
		\label{fig:non_seq_scheme}
	\end{minipage}
	\hspace{0.5cm}
	\begin{minipage}[c]{0.65\textwidth}
		\centering
		\includegraphics[width=0.9\textwidth]{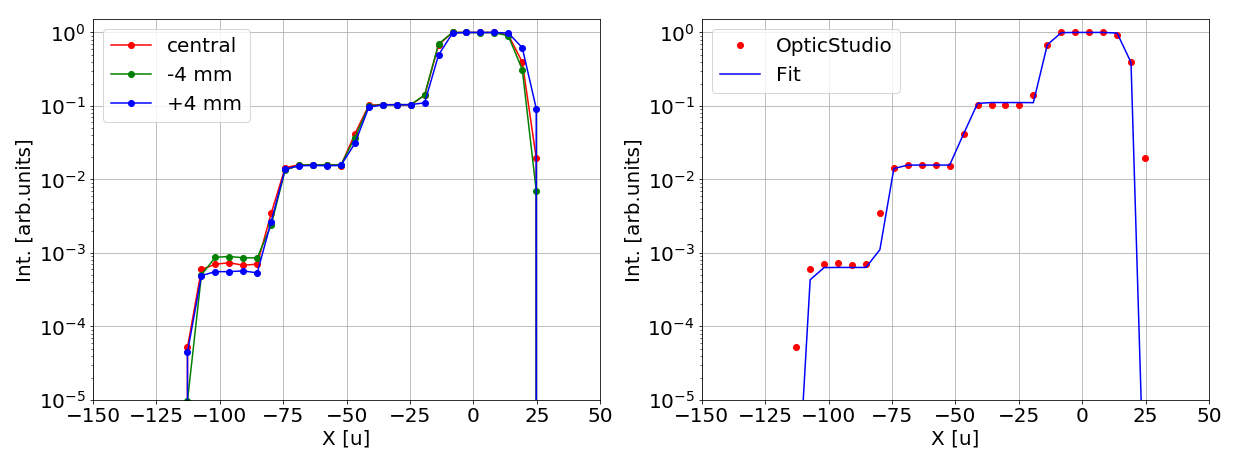}
		\caption{Results of $PSF_{scint}$ simulation: left - comparison of on--axis and off--axis PSFs; right - 		fitting of the central one.}
		\label{fig:psf_scint}
	\end{minipage}
\end{figure}

Following the analysis of $PSF_{aber}$, attention shifts to $PSF_{scint}$. First, a few words about scheme of simulation. The geometry is the same as for $PSF_{aber}$. The lens in this case is a thin lens because the aberrations were already taken into account. In the non-sequential mode of OpticStudio\textsuperscript{\textcopyright} the scintillator bulk was modeled with a line of point sources put in it - see Fig.~\ref{fig:non_seq_scheme}. Each point source radiates isotropically in $4\pi$. In the figure one may see four rays which correspond to four light sources. In the simulation there were 20. Further increment of the sources number didn't affect the result. The resulting PSFs are presented in Fig.~\ref{fig:psf_scint}. Similar to the aberrational PSF analysis, this segment also compares off-axis and on-axis PSFs. The three lower steps are attributed to internal reflections within the system. For example the second step represents the rays shown with dashed arrows in Fig.~\ref{fig:non_seq_scheme}. The last step exhibits some deviation between off and on-axis PSFs. Nevertheless, this discrepancy is negligible for our purposes: it is three orders of magnitude lower than the principal component and thus exerts an insubstantial influence on the overall image quality. The fit of the central PSF is shown on the right-hand plot. It was fitted with a function, composed of four higher-order Gaussians:
\begin{equation}
	\begin{gathered}
	\label{eq:psf_scint}
	PSF_{scint} = A_1 \exp\left[ \frac{- (x - \mu)^{10}}{2 \sigma^{10}} \right] +
				  A_2 \exp\left[ \frac{- (x - \mu - \Delta\mu)^{10}}{2 \sigma^{10}} \right] + \\
				  A_3 \exp\left[ \frac{- (x - \mu - 2\Delta\mu)^{10}}{2 \sigma^{10}} \right] +
				  A_4 \exp\left[ \frac{- (x - \mu - 3\Delta\mu)^{10}}{2 \sigma^{10}} \right] +
	\end{gathered}.
\end{equation}

\begin{figure}[!hbt]
	\centering
	\includegraphics[width=0.65\textwidth]{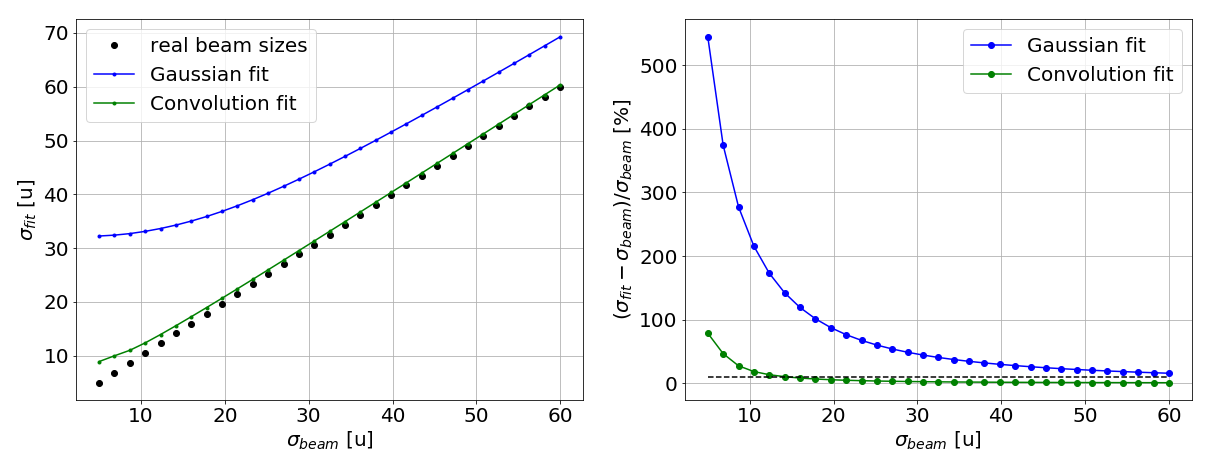}
	\caption{Fit results: left - comparison of the fitted and the actual sizes; right - deviation from the 			real sizes.}
	\label{fig:fit_res}
\end{figure}

With the PSFs at hand, simulating an image of a gaussian beam with Eq.~\ref{eq:image_formation3} becomes feasible. Then by fitting images representing various beam sizes, we can deduce the station's resolution. The natural choice for the fitting function is the Gaussian. However, given our knowledge of both PSFs, we opt to enhance our fitting procedure by also using the convolution of the Gaussian with the PSFs. The outcomes of these fits are depicted in Fig.~\ref{fig:fit_res}. In the left figure, the black dots represent the actual beam sizes, while the blue and green curves correspond to the fits using the Gaussian function and its convolution with the PSFs, respectively. The right figure quantifies the deviation from the actual beam sizes, with the dashed black line indicating a tolerance level of $10\%$. It is observed that the deviation for the Gaussian fit stabilizes around a beam size of $50$ $\mu m$ and only crosses the $10\%$ threshold at $60$ $\mu m$. Conversely, the fit incorporating the convolved Gaussian demonstrates significant deviations for beam sizes smaller than $20$ $\mu m$ and exceeds the $10\%$ level below $10$ $\mu m$. Thus, the station's resolution is approximated to be around $32$ $\mu m$.

\begin{figure}[!hbt]
	\centering
	\includegraphics[width=0.7\textwidth]{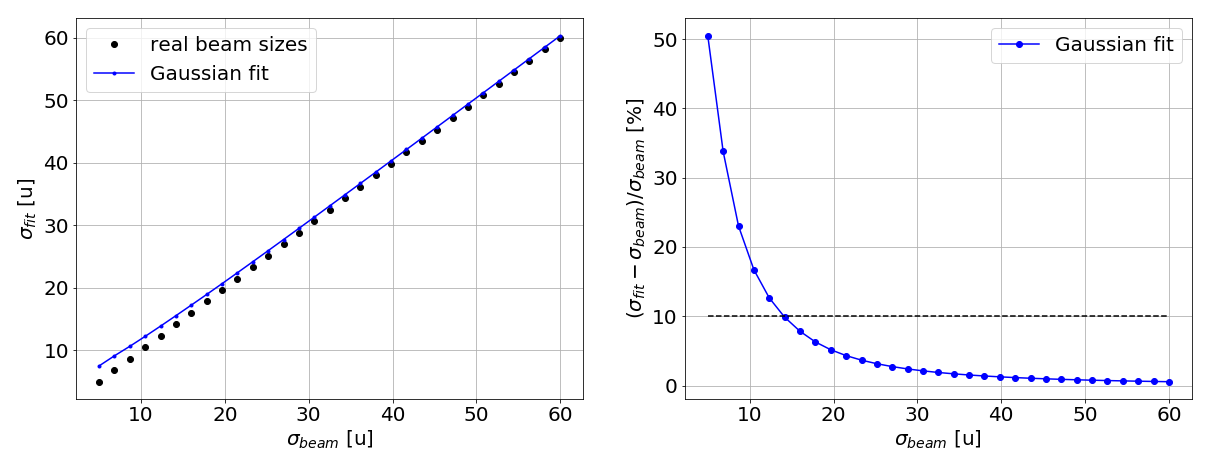}
	\caption{Fit results: comparison of the fitted and the actual sizes; right - deviation from the real sizes.}
	\label{fig:fit_res_no_angle_plane}
\end{figure}

It is intriguing to contrast these results with those obtained from the plane not affected by the observation angle. The outcomes for this scenario are presented in Fig.~\ref{fig:fit_res_no_angle_plane}. Here, the fit was performed solely using a Gaussian function. The estimated resolution in this case is around $8$ $\mu m$. Since this plane is influenced only by $PSF_{aber}$, which exhibits Gaussian characteristics with $\sigma_{aber}=2.18\, \mu m$, we can straightforwardly adjust for it using the well--known formula $\sigma^2 = \sigma_{fit}^2 - \sigma_{aber}^2$.

\subsection*{1:1 optics}
\begin{figure}[!hbt]
	\centering
	\includegraphics[width=0.7\textwidth]{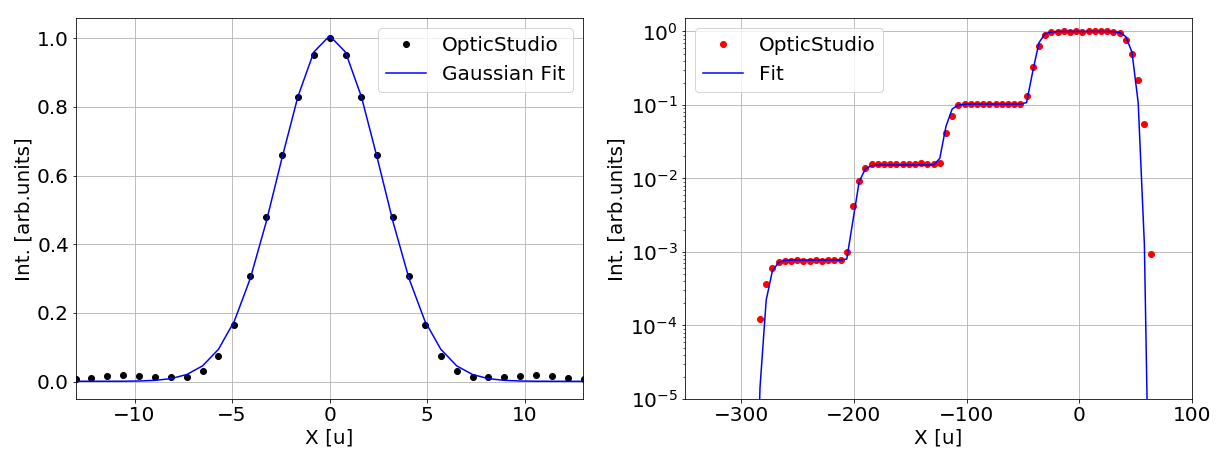}
	\caption{$PSF_{aber}$ and $PSF_{scint}$.}
	\label{fig:OTRA_psf_aber_scint}
\end{figure}

The same simulation was also conducted for the $1:1$ optics setup, which utilizes a different configuration compared to the $1:2$ setup. This includes a different objective, the \textit{Schneider-Kreuznach 5.6/180}, and a modified geometry. The distances from the scintillator to the objective and from the objective to the camera are both $360$ $mm$, with the camera tilted at $45^\circ$. The scintillator, camera, and other parameters remain unchanged. In Fig.~\ref{fig:OTRA_psf_aber_scint}, both $PSF_{aber}$ and $PSF_{scint}$ are displayed. The aberration PSF on the left-hand side is fitted with a Gaussian function characterized by $\sigma = 2.6\, \mu m$, which is slightly wider than that observed in the $1:2$ geometry due to the different objective used.

\begin{figure}[!hbt]
	\centering
	\includegraphics[width=0.7\textwidth]{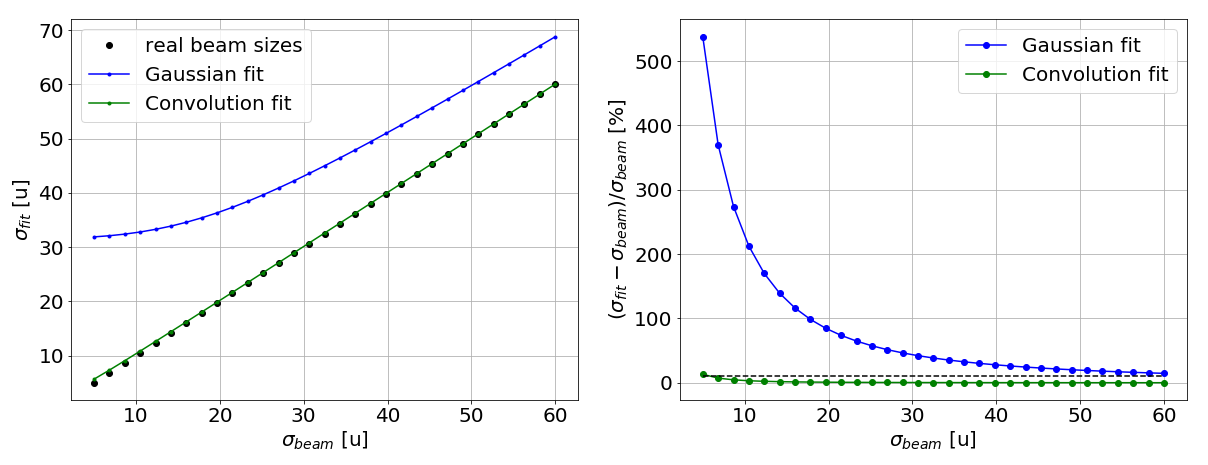}
	\caption{Fit results: left - comparison of the fitted and the real sizes; right - deviation from the real sizes.}
	\label{fig:fit_res_otra}
\end{figure}

The beam sizes ranging from 5 to 60 $\mu m$ have been simulated. The results are depicted in Fig.~\ref{fig:fit_res_otra}. The resolution determined through the standard Gaussian fit approximates 32 $\mu m$. Notably, the result from the convolved Gaussian fit exhibits even better accuracy, only deviating beyond the $10 \, \%$ level of discrepancy at a beam size of 5 $\mu m$.

\begin{figure}[!hbt]
	\centering
	\includegraphics[width=0.7\textwidth]{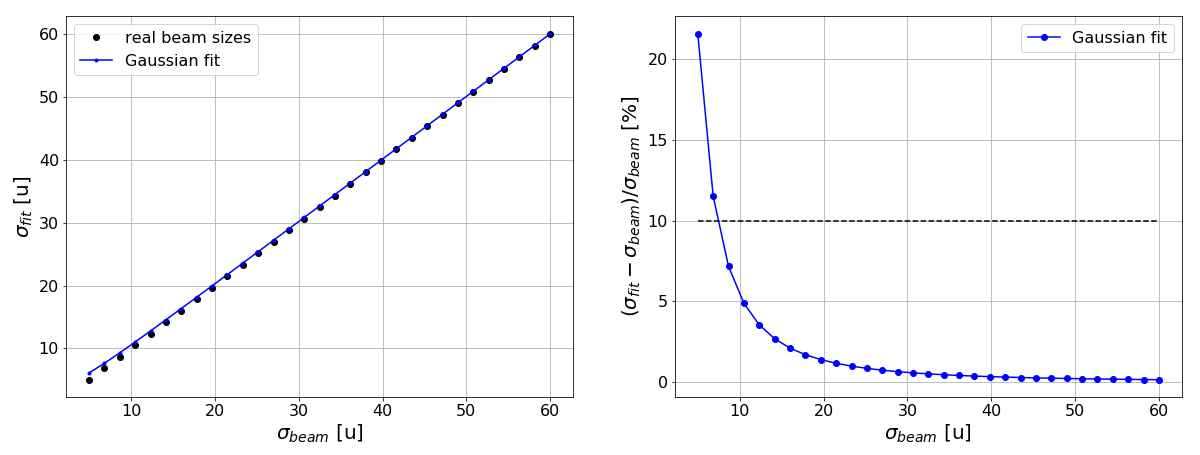}
	\caption{Fit results of the plane not affected by the observation angle.}
	\label{fig:otra_fit_res_no_angle}
\end{figure}

The results for the other plane are illustrated in Fig.~\ref{fig:otra_fit_res_no_angle}, revealing a resolution of 6 $\mu m$, which is remarkably smaller than that achieved with the $1:2$ optics. This finding may seem counterintuitive since the $PSF_{aber}$ is actually smaller in the $1:2$ case.
Lets take a closer look at our simulation. We model a gaussian beam with a $\sigma_{beam}$ width. The $PSF_{aber}$ is also represented by a gaussian with $\sigma_{aber}$ width. Additionally, the optical system's magnification $M$ is either 1 or 0.5. consequently, the width of the image beam in the plane unaffected by the observation angle is defined as follows:
\begin{equation}
	\label{eq:img_sigma}
	\sigma_{image}^2 = (M \sigma_{beam})^2 + \sigma_{aber}^2.
\end{equation}
In the setup only the beam's sigma is altered by the magnification, since aberrations depend solely on the objective lens. However, during the image processing, we apply the magnification factor to entire image ($\sigma_{image}$), affecting both $\sigma_{beam}$ and $\sigma_{aber}$. For the $1:1$ optical system there is no effect, but with the $1:2$ optics the aberrations are artificially magnified, leading to an enlargement in the final results.

\section*{Experimental approval of the models}
\begin{figure}[!hbt]
	\centering
	\subfigure[5.6/180 objective.]{
		\includegraphics[width=0.4\textwidth]{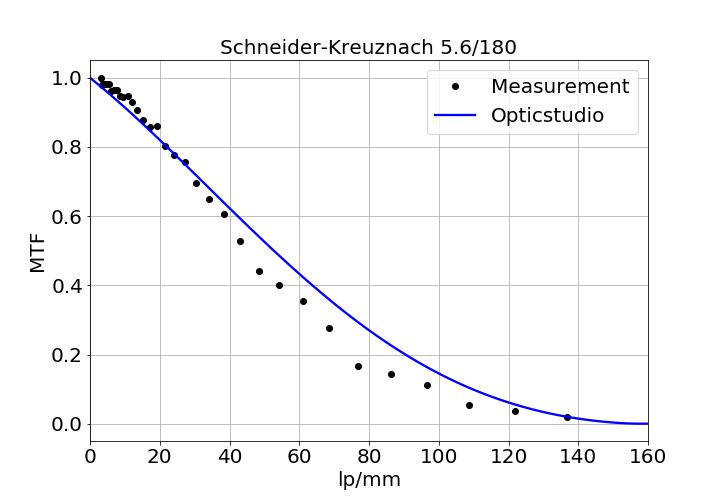}
		\label{fig:mtf_res_56_180}
	}
	\subfigure[5.9/120 objective.]{
		\includegraphics[width=0.4\textwidth]{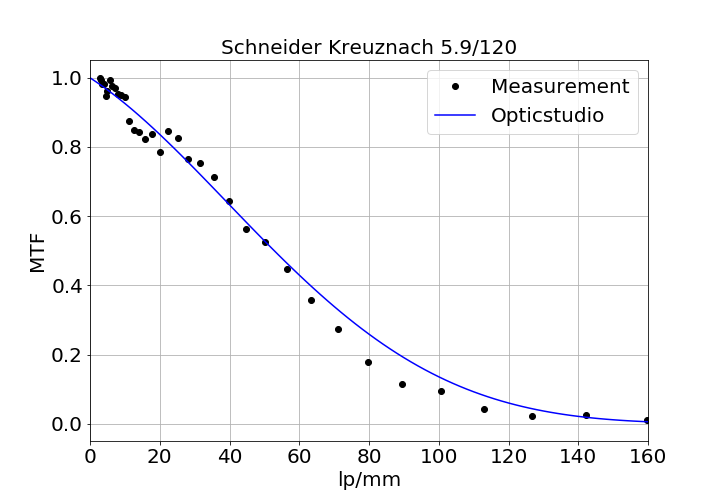}
		\label{fig:mtf_res_59_120}
	}
	\caption{Measured and modeled MTF results.}
	\label{fig:mtf_res}
\end{figure}

The models have to be approved. First the $PSF_{aber}$ approval will be demonstrated, which was done by a Modulation Transfer Function (MTF) measurement and comparison to results from OpticStudio\textsuperscript{\textcopyright}. To explain a bit about the MTF we need to start from the Fourier transformation of the $PSF_{aber}$, the result of which is an Optical Transfer Function (OTF):
\begin{equation}
	\mathcal{F}\{PSF(x)\} = OTF(\nu).
\end{equation}
The OTF, in turn, can be expressed in terms of the MTF and the Phase Transfer Function (PhTF):
\begin{equation}
	OTF(\nu) = MTF(\nu)\, e^{PhTF(\nu)}.
\end{equation}
We will not delve deeply into the details of transfer functions, as they are extensively covered in many textbooks, such as in \cite{Optics}, just some general words. The OTF describes the resolving capability of an optical system. The MTF quantifies how well the system can transfer various levels of detail, contrast at different spatial frequencies in particular. Lastly, the PhTF shows how the optical system affects the phase of the spatial frequency components. In our approach we used only the MTF. To measure it an appropriate target like the USAF 1951 target, a star target, or a sinusoidal target \cite{test_targets} is required. We have used the USAF 1951 target. Skipping the details of the target and its processing, as it may be found in many textbooks, we will focus on the results. Fig.~\ref{fig:mtf_res} shows the measurement results (black dots) and the modeling results (blue lines) for both \textit{5.6/180} (left) and \textit{5.9/120} (right) objectives. It is worth mentioning that the narrower the PSF, the wider the MTF. From the graphs, we can observe that the measurements show slightly worse results in the modulation frequency range of $50 \div 120$. Although the disagreement is not significant, let's examine the difference in terms of PSF if we fit the MTF to the measurement results.

\begin{figure}[hbt!]
	\centering
	\subfigure[MTF fitting.]{
		\includegraphics[width=0.4\textwidth]{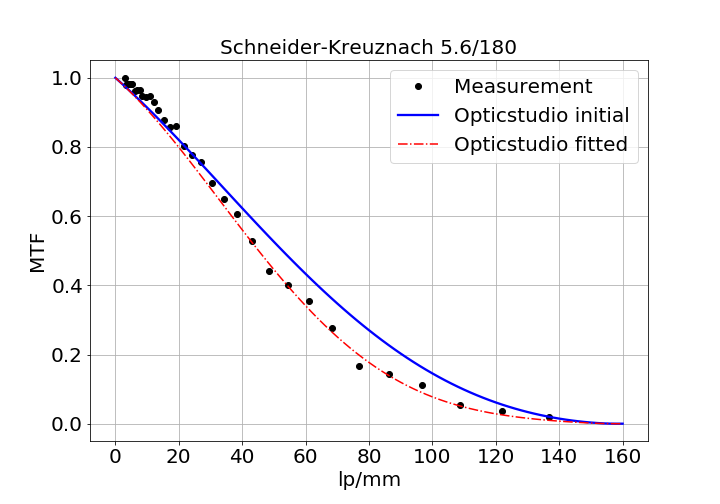}
		\label{fig:mtf_fit}
	}
	\subfigure[Corresponding PSFs.]{
		\includegraphics[width=0.4\textwidth]{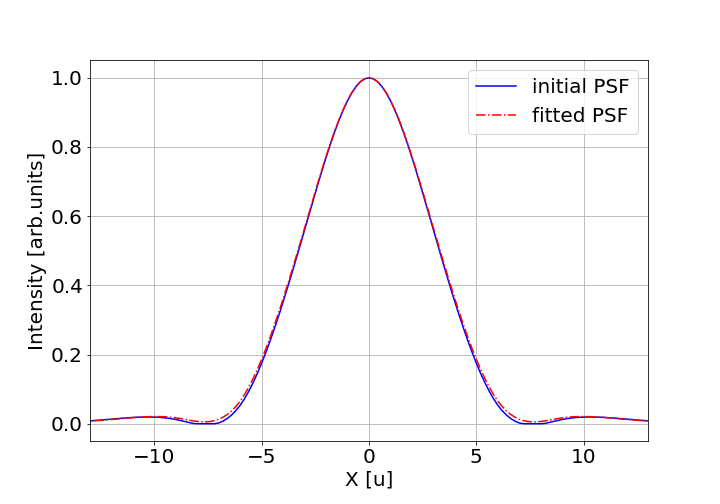}
		\label{fit:mtf_fit_psf}
	}
	\caption{Initial and fitted MTFs and corresponding PSFs.}
	\label{fig:mtf_psf_fit_res}
\end{figure}

In order to do that, the image plane in OpticStudio was slightly displaced until it matched the measured data. This adjustment led to a widening (defocusing in fact) of the PSF. The results are shown in Fig.~\ref{fig:mtf_psf_fit_res}. The blue lines represent the original OpticStudio\textsuperscript{\textcopyright} MTF for the \textit{5.6/180} objective, while the red line represents the MTF fitted to the measured data. The right-hand figure displays the corresponding PSFs. There is only a very slight disagreement in the region of $\pm 5 \div \pm 10$ $\mu m$, which is undetectable in our optical system. Referring back to Fig.~\ref{fig:mtf_res_59_120}, the \textit{5.9/120} objective shows even less discrepancy, suggesting that the PSFs differ even less from the modeled ones. Thus, the $PSF_{aber}$ part is validated.

Now the $PSF_{scint}$ needs to be validated. The ideal way to do this would be to image an electron beam with a transverse size significantly smaller than $PSF_{aber}$ as it passes through the scintillator. However, creating such a beam at the XFEL or other machines at DESY is a challenging task. Nevertheless, proving the overall station resolution is sufficient, given that $PSF_{aber}$ has already been validated. Additionally, it is only necessary to do this for either the $1:1$ or $1:2$ station type, as this part of the resolution is independent of the objective. This validation has been carried out during the measurements described in \cite{ffw_resolution, tomin_zagorodnov}. The work \cite{ffw_resolution} reports a resolution of around $6 \, \mu m$ in the plane unaffected by $PSF_{scint}$ for the $1:1$ station type, which coincides to our calculated value of $6.2 \, \mu m$. The studies \cite{tomin_zagorodnov} estimate the resolution in the plane affected by $PSF_{scint}$ to be around $28 \, \mu m$ for the $1:1$ station type, which is $4 \, \mu m$ off from our model, but still reasonably close.

\section*{Conclusion}
The two main geometries used at the European XFEL have been modeled in OpticStudio and validated experimentally. This validation allows us to confidently use these models for any possible geometries, provided the same objectives are used, thereby generally improving the monitor's resolution.

Both $PSF_{scint}$ and $PSF_{aber}$ can be included in the fitting procedure, significantly enhancing the resolution for beam sizes smaller than $50 \, \mu m$. The $PSF_{scint}$ shown here is obviously only applicable to the $200 \, \mu m$ scintillator, but any other scintillator thickness may easily be simulated and fitted with the Eq.~\ref{eq:psf_scint}.

This method offers an alternative to physical modifications of the setup to improve resolution, particularly when such modifications are technically challenging, such as using a thinner scintillator.


\end{document}